\documentclass[envcountsame]{llncs}

\usepackage{amssymb}
\title{{\bf On Infinitary Rational Relations \\ and Borel Sets} }

\author{Olivier Finkel\inst{}}
\institute{Equipe de Logique Math\'ematique \\
 U.F.R. de Math\'ematiques, Universit\'e Paris 7 \\ 2 Place Jussieu 75251 Paris
 cedex 05, France \\ \email{finkel@logique.jussieu.fr}.}

\date{}

\begin{document}

\spnewtheorem{Rem}[theorem]{Remark}{\bfseries}{\itshape}
\spnewtheorem{Exa}[theorem]{Example}{\bfseries}{\itshape}

\spnewtheorem{Pro}[theorem]{Proposition}{\bfseries}{\itshape}
\spnewtheorem{Lem}[theorem]{Lemma}{\bfseries}{\itshape}
\spnewtheorem{Cor}[theorem]{Corollary}{\bfseries}{\itshape}

\newcommand{\fa}{\forall}
\newcommand{\Ga}{\Gamma}
\newcommand{\Gas}{\Gamma^\star}
\newcommand{\Si}{\Sigma}
\newcommand{\Sis}{\Sigma^\star}
\newcommand{\Sio}{\Sigma^\omega}
\newcommand{\ra}{\rightarrow}
\newcommand{\hs}{\hspace{12mm}

\noi}
\newcommand{\lra}{\leftrightarrow}
\newcommand{\la}{language}
\newcommand{\ite}{\item}
\newcommand{\Lp}{L(\varphi)}
\newcommand{\abs}{\{a, b\}^\star}
\newcommand{\abcs}{\{a, b, c \}^\star}
\newcommand{\ol}{ $\omega$-language}
\newcommand{\orl}{ $\omega$-regular language}
\newcommand{\om}{\omega}
\newcommand{\nl}{\newline}
\newcommand{\noi}{\noindent}
\newcommand{\tla}{\twoheadleftarrow}
\newcommand{\de}{deterministic }
\newcommand{\proo}{\noi {\bf Proof.} }
\newcommand {\ep}{\hfill $\square$}

\maketitle

\begin{abstract}
\noi We prove in this paper that there exists some infinitary rational relations 
which are ${\bf \Si^0_3}$-complete 
Borel sets and some others which are ${\bf \Pi^0_3}$-complete. 
These results give additional  answers to 
 questions of Simonnet \cite{sim} and of Lescow and Thomas \cite{tho} \cite{lt}. 
\end{abstract}

\noi {\small {\bf Keywords:} infinitary rational relations; topological properties; Borel sets.}

\section{Introduction}

Acceptance of infinite words by finite automata was firstly 
considered in the sixties by B\"uchi in order to study 
decidability of the monadic second order theory 
of one successor over the integers \cite{bu62}. Then the so called \orl s have been 
intensively studied and many applications have been found, see \cite{tho} \cite{sta} 
\cite{pp} for many results and references. 
\nl Since then many extensions of \orl s have been investigated as the classes of 
\ol s accepted by pushdown automata, Petri nets, Turing machines, 
see \cite{tho} \cite{eh} \cite{sta} for a survey of this work. 
\nl On the other side
rational relations on finite words were studied in the sixties and played 
a fundamental role in the study of families of context free languages \cite{ber}. 
Investigations on their extension to rational  relations on infinite words were carried out 
or mentioned in the books \cite{bt} \cite{ls}. Gire  and Nivat 
studied infinitary rational relations in  \cite{gire1} \cite{gn}. 
Infinitary rational relations 
are subsets of  $\Si_1^\om \times \Si_2^\om$,  where 
$\Si_1$ and  $\Si_2$ are finite alphabets, which are recognized by 
B\"uchi transducers or by $2$-tape finite B\"uchi automata with asynchronous 
reading heads. 
So the class of infinitary rational relations extends both the 
class of finitary rational relations {\bf and } the class of \orl s.  
\nl They have been much studied, in particular in connection with 
the rational functions they may define, see for example  \cite{cg} \cite{bcps} 
\cite{sim} \cite{sta} \cite{pri} for many results and references. Notice that 
a rational relation $R\subseteq \Si_1^\om \times \Si_2^\om$ may be seen as 
an \ol~ over the alphabet $\Si_1 \times \Si_2$. 
\nl  A way to study the complexity of languages of infinite words  
accepted by finite machines is to study their topological complexity and firstly
to locate them with regard to 
the Borel and the projective hierarchies. 
This work is analysed 
for example in \cite{stac} \cite{tho} \cite{eh} \cite{lt} \cite{sta}.  
It is well known that every \ol~ accepted by a Turing machine with a 
B\"uchi or Muller acceptance condition is an analytic set and 
 that \orl s are boolean combinations of ${\bf \Pi_2^0}$-sets 
hence ${\bf \Delta_3^0}$-sets,  \cite{sta} \cite{pp}.  
\nl  The question of the topological complexity of  relations on infinite words also 
naturally arises and is asked by Simonnet in \cite{sim}. It is also posed in a more 
general form by Lescow and Thomas in \cite{lt} 
(for infinite labelled partial orders) and in \cite{tho89} 
where Thomas suggested to study reducibility notions and associated completeness results.  
\nl Every infinitary rational relation is an analytic set. 
 We showed in \cite{relrat} that there exist some infinitary rational relations 
which are analytic but non  Borel sets. 
 Considering Borel infinitary rational relations we prove in this paper 
that there exist some infinitary rational relations 
which are ${\bf \Si^0_3}$-complete 
Borel sets and some others which are ${\bf \Pi^0_3}$-complete. 
This implies that there exist also some infinitary rational relations 
which are ${\bf \Delta^0_4}$-sets but not  (${\bf \Si^0_3} \cup{\bf \Pi^0_3}$)-sets. 
\nl These results may be compared with examples of ${\bf \Si^0_3}$-complete \ol s accepted 
by deterministic pushdown automata with the acceptance condition: ``some stack content 
 appears infinitely often during an infinite run",   
given by Cachat, Duparc,  and Thomas in \cite{cdt} 
or with examples 
of ${\bf \Si^0_n}$-complete and ${\bf \Pi^0_n}$-complete  \ol s,  $n\geq 1$,  accepted  by  
non-deterministic pushdown automata with B\"uchi acceptance condition given in \cite{fina}. 
\nl  The paper is organized as follows. In section 2 we introduce 
the notion of transducers and 
of infinitary rational relations. In section 3 we recall definitions of Borel
sets, and we prove our main results  in sections 4 and 5.

\section{Infinitary rational relations}

\noi  Let $\Si$ be a finite alphabet whose elements are called letters.
A non-empty finite word over $\Si$ is a finite sequence of letters:
 $x=a_1a_2\ldots a_n$ where for all $i$ in $[1; n]$,  $a_i \in\Si$.
 We shall denote $x(i)=a_i$ the $i^{th}$ letter of $x$
and $x[i]=x(1)\ldots x(i)$ for $i\leq n$. The length of $x$ is $|x|=n$.
The empty word will be denoted by $\lambda$ and has 0 letter. Its length is 0.
 The set of finite words over $\Si$ is denoted $\Sis$.
 A (finitary) language $L$ over $\Si$ is a subset of $\Sis$.
 The usual concatenation product of $u$ and $v$ will be denoted by $u.v$ or just  $uv$.
 For $V\subseteq \Sis$, we denote  \quad 
$V^\star = \{ v_1\ldots v_n  \mid 
  n\in \mathbb{N}\quad and \quad \fa i \in [1; n] \quad  v_i \in V \}$.

\hs The first infinite ordinal is $\om$.
An $\om$-word over $\Si$ is an $\om$ -sequence $a_1a_2\ldots a_n \ldots$, where 
for all $i \geq 1$,  
$a_i \in\Sigma$. 
 When $\sigma$ is an $\om$-word over $\Si$, we write
 $\sigma =\sigma(1)\sigma(2)\ldots  \sigma(n) \ldots $
and $\sigma[n]=\sigma(1)\sigma(2)\ldots  \sigma(n)$ the finite word of length $n$, 
prefix of $\sigma$.
The set of $\om$-words over  the alphabet $\Si$ is denoted by $\Si^\om$.
 An  $\om$-language over an alphabet $\Sigma$ is a subset of  $\Si^\om$.
For $V\subseteq \Sis$,
 $V^\om = \{ \sigma =u_1\ldots  u_n\ldots  \in \Si^\om \mid  \fa i\geq 1 ~ u_i\in V\}$
is the $\om$-power of $V$.
 The concatenation product is extended to the product of a 
finite word $u$ and an $\om$-word $v$: 
the infinite word $u.v$ is then the $\om$-word such that:
 $(u.v)(k)=u(k)$  if $k\leq |u|$ , and  $(u.v)(k)=v(k-|u|)$  if $k>|u|$.

\hs If $A$ is a subset of $B$ we shall denote $A^-=B-A$ the complement of $A$ (in $B$). 

\hs We assume the reader to be familiar with the theory of formal languages and of 
\orl s. We recall that \orl s form the class of \ol s accepted 
by finite automata with a   B\"uchi acceptance condition and this class is the omega Kleene 
closure of the class of regular finitary languages. 
\nl We are going now to introduce  the notion of infinitary rational relation
via definition of  B\"uchi transducers:

\begin{definition}
A B\"uchi transducer is a sextuple $\mathcal{T}=(K, \Si, \Ga, \Delta, q_0, F)$, where 
$K$ is a finite set of states, $\Si$ and $\Ga$ are finite sets called the input and 
the output alphabets, 
$\Delta$ is a finite subset of $K \times \Sis \times \Gas \times K$ called 
the set of transitions, $q_0$ is the initial state,  and $F \subseteq K$ is the set of 
accepting states. 
\nl A computation $\mathcal{C}$ of the transducer $\mathcal{T}$ is an infinite sequence of transitions 
$$(q_0, u_1, v_1, q_1), (q_1, u_2, v_2, q_2), \ldots (q_{i-1}, u_{i}, v_{i}, q_{i}), 
(q_i, u_{i+1}, v_{i+1}, q_{i+1}), \ldots $$
\noi The computation is said to be successful iff there exists a final state $q_f \in F$ 
and infinitely many integers $i\geq 0$ such that $q_i=q_f$. 
\nl The input word of the computation is $u=u_1.u_2.u_3 \ldots$
\nl The output word of the computation is $v=v_1.v_2.v_3 \ldots$
\nl Then the input and the output words may be finite or infinite. 
\nl The infinitary rational relation $R(\mathcal{T})\subseteq \Sio \times \Ga^\om$ 
recognized by the B\"uchi transducer $\mathcal{T}$ 
is the set of pairs $(u, v) \in \Sio \times \Ga^\om$ such that $u$ and $v$ are the input 
and the output words of some successful computation $\mathcal{C}$ of $\mathcal{T}$. 
\nl The set of infinitary rational relations will be denoted $RAT$. 
\end{definition}

\begin{Rem}\label{ol s}
An infinitary rational relation is a subset of $\Sio \times \Ga^\om$ for two finite 
alphabets $\Si$ and $\Ga$. One can also consider that it is an \ol~ over the finite 
alphabet $\Si \times \Ga$. If $(u, v) \in \Sio \times \Ga^\om$,
 one can consider this pair of 
infinite words as a single infinite word $(u(1),v(1)).(u(2), v(2)).(u(3), v(3))\ldots $
over the alphabet $\Si \times \Ga$. 
 We shall use this fact to investigate the topological complexity 
of infinitary rational relations. 
 \end{Rem}

\section{Borel sets}

\noi We assume the reader to be familiar with basic notions of topology which
may be found in  \cite{mos} \cite{kec} \cite{lt} \cite{sta} \cite{pp}.

\hs For a  finite alphabet $X$ 
 we shall consider $X^\om$ as a topological space with the Cantor topology.
 The open sets of $X^\om$ are the sets in the form $W.X^\om$, where $W\subseteq X^\star$.
A set $L\subseteq X^\om$ is a closed set iff its complement $X^\om - L$ is an open set.
\nl  Define now the next classes of the  Hierarchy of Borel sets of finite ranks:

\begin{definition}
The classes ${\bf \Si_n^0}$ and ${\bf \Pi_n^0 }$ of the Borel Hierarchy
 on the topological space $X^\om$  are defined as follows:
\nl ${\bf \Si^0_1 }$ is the class of open sets of $X^\om$.
\nl ${\bf \Pi^0_1 }$ is the class of closed sets of $X^\om$.
\nl And for any integer $n\geq 1$:
\nl ${\bf \Si^0_{n+1} }$   is the class of countable unions 
of ${\bf \Pi^0_n }$-subsets of  $X^\om$.
\nl ${\bf \Pi^0_{n+1} }$ is the class of countable intersections of 
${\bf \Si^0_n}$-subsets of $X^\om$.

\end{definition}

\noi  The Borel Hierarchy is also defined for transfinite levels, but we shall not 
need them in the present study. 
There are also some subsets of $X^\om$ which are not Borel.  In particular 
the class of Borel subsets of $X^\om$ is strictly included into 
the class  ${\bf \Si^1_1}$ of analytic sets which are 
obtained by projection of Borel sets, 
see for example \cite{sta} \cite{lt}  \cite{pp} \cite{kec} for more details.

\hs Recall also the notion of completeness with regard to reduction by continuous functions. 
For an integer $n\geq 1$, a set $F\subseteq X^\om$ is said to be 
a ${\bf \Si^0_n}$  (respectively,  ${\bf \Pi^0_n}$, ${\bf \Si^1_1}$)-complete set 
iff for any set $E\subseteq Y^\om$  (with $Y$ a finite alphabet): 
 $E\in {\bf \Si^0_n}$ (respectively,  $E\in {\bf \Pi^0_n}$,  $E\in {\bf \Si^1_1}$) 
iff there exists a continuous function $f: Y^\om \ra X^\om$ such that $E = f^{-1}(F)$.  
\nl  A ${\bf \Si^0_n}$
 (respectively,  ${\bf \Pi^0_n}$, ${\bf \Si^1_1}$)-complete set is a ${\bf \Si^0_n}$
 (respectively,  ${\bf \Pi^0_n}$, ${\bf \Si^1_1}$)-set which is 
in some sense a set of the highest 
topological complexity among the ${\bf \Si^0_n}$
 (respectively,  ${\bf \Pi^0_n}$, ${\bf \Si^1_1}$)-sets. 
 ${\bf \Si^0_n}$
 (respectively,  ${\bf \Pi^0_n}$)-complete sets, with $n$ an integer $\geq 1$, 
 are thoroughly characterized in \cite{stac}.  

  \begin{Exa}\label{exa} Let $\Si=\{0, 1\}$ and $\mathcal{A}=(0^\star.1)^\om \subseteq \Sio$. 
$\mathcal{A}$ is the set of 
$\om$-words over the alphabet $\Si$ with infinitely many occurrences of the letter $1$. 
It is well known that $\mathcal{A}$ is a 
${\bf \Pi^0_2 }$-complete set and its complement $\mathcal{A}^-$ 
is a ${\bf \Si^0_2 }$-complete set: it is the set of $\om$-words over $\{0, 1\}$ having 
only a finite number of occurrences of letter $1$. 
\end{Exa}

\section{${\bf \Si^0_3}$-complete infinitary  rational relations}

\noi We can now state the following result:

\begin{theorem}\label{si30} There exist some ${\bf \Si^0_3}$-complete 
infinitary  rational relations. 
\end{theorem}

\proo   We shall use a well known example of ${\bf \Si^0_3}$-complete set which is a 
subset of the topological space $\Si^{\om^2}$. 

\hs The set  $\Si^{\om^2}$ is the set of $\om^2$-words over the finite alphabet $\Si$. 
It may also be viewed as the set of (infinite) $(\om \times \om)$-matrices 
whose coefficients are letters 
of $\Si$. If $x \in \Si^{\om^2}$ we shall write $x = (x(m, n))_{m\geq 1, n\geq 1}$.  
The infinite word $x(m, 1)x(m, 2)\ldots x(m, n)\ldots$ will be called 
the $m^{th}$ column of the $\om^2$-word $x$ and the infinite word 
$x(1, n)x(2, n)\ldots x(m, n)\ldots$ will be called 
the $n^{th}$ row of the $\om^2$-word $x$.  Thus an element of   $\Si^{\om^2}$ 
is completely determined by the (infinite) set of its columns or of its rows. 
\nl The set $\Si^{\om^2}$ is usually equipped with the product topology  of the dicrete 
topology on $\Si$ (for which every subset of $\Si$ is an open set), see \cite{kec} \cite{pp}. 
This topology may be defined 
by the following distance $d$. Let $x$ and $y$ be two  $\om^2$-words in $\Si^{\om^2}$ 
such that $x\neq y$, then  
$$ d(x, y)=\frac{1}{2^n} ~~~~~~~\mbox{   where  }$$
$$n=min\{p\geq 1 \mid  \exists (i, j) ~~ x(i, j)\neq y(i, j) \mbox{ and } i+j=p\}$$

\hs Then the topological space $\Si^{\om^2}$ is homeomorphic to the 
topological space  $\Si^{\om}$, equipped with the Cantor topology.  
 Borel  subsets  of   $\Si^{\om^2}$ are defined from open 
subsets in the same manner as in the case of the topological space $\Si^\om$. 
\nl  ${\bf \Si^0_n}$ (respectively ${\bf \Pi^0_n}$)-complete sets are also 
defined in a similar way.  

\hs Recall now that the set 
$$S = \{ x\in \{0, 1\}^{\om^2}~/~ \exists m \exists^\infty n ~x(m, n)=1\}$$ 
\noi where $\exists^\infty$ means "there exist infinitely many", 
is a ${\bf \Si^0_3}$-complete subset of $\{0, 1\}^{\om^2}$, \cite[p. 179]{kec}. 
It is the set of $\om^2$-words over $\{0, 1\}$ 
having at least one column in the ${\bf \Pi^0_2 }$-complete subset 
$\mathcal{A}$ of $\{0, 1\}^{\om}$ 
given in Example \ref{exa}. 

\hs In order to use this example we shall firstly define a coding of $\om^2$-words 
over $\Si$ by $\om$-words over the alphabet  $(\Si\cup\{A\}) \times (\Si\cup\{A\})$ where  
$A$ is a new letter not in $\Si$.
The code of $x\in \Si^{\om^2}$ may be written in the form $(\sigma_1, \sigma_2)$, 
where $\sigma_1$ and $\sigma_2$ are 
$\om$-words over the alphabet $(\Si\cup\{A\})$.  In order to describe 
the $\om$-words  $\sigma_1$ and $\sigma_2$ let us call, 
for $x\in \Si^{\om^2}$ and $p$ an integer $\geq 2$:  
$$T^x_{p+1}=\{x(p,1), x(p-1, 2), \ldots , x(2, p-1), x(1,p)\}$$ 
the set of elements $x(m, n)$ with $m+n=p+1$. 
\nl The word $\sigma_1$ begins with $x(1, 1)$ followed by a letter $A$; then the 
word $\sigma_1$ enumerates the elements of the sets $T^x_{p+1}$ for $p$ an {\bf even integer}. 
More precisely for every even integer $p\geq 2$ the elements of $T^x_{p+1}$ are placed before 
those of $T^x_{p+3}$ and these two sets of letters are separated by an $A$. Moreover for each 
even integer $p\geq 2$ the elements of $T^x_{p+1}$ are enumerated in the following order: 

$$ x(p,1), x(p-1, 2), \ldots , x(2, p-1), x(1,p)$$

\noi The construction of $\sigma_2$ is very similar but with some modifications. It 
enumerates the elements of the sets $T^x_{p+1}$ for $p$ an {\bf odd integer}. More precisely 
the word  $\sigma_2$ begins with an $A$ then for every odd integer $p\geq 3$ 
the elements of $T^x_{p+1}$ are placed before 
those of $T^x_{p+3}$ and these two sets of letters are separated by an $A$. Moreover for each 
odd integer $p\geq 3$ the elements of $T^x_{p+1}$  are enumerated in the following order:

$$ x(1,p), x(2, p-1),  \ldots, x(p-1, 2),  x(p,1)$$

\noi Then the $\om$-word $\sigma_1$ and  the $\om$-word $\sigma_2$ are in the following form:

\hs $\sigma_1=x(1,1).A.x(2,1)x(1,2).A.x(4,1)x(3,2)x(2,3)x(1,4).A.x(6,1)x(5,2)\ldots $

\hs $\sigma_2=A.x(1,3)x(2,2)x(3,1).A.x(1,5)x(2,4)x(3,3)x(4,2)x(5,1).A.x(1,7)x(2,6)\ldots $

\hs Let then $h$ be the mapping from $\Si^{\om^2}$  into
 $((\Si\cup\{A\})\times (\Si \cup\{A\}))^\om$ 
such that,  for every $\om^2$-word $x$ over the  alphabet $\Si$,  
$h(x)$ is the code $(\sigma_1, \sigma_2)$ of the $\om^2$-word  as defined above.
It is easy to see, from the definition of $h$ and of the  order of the enumeration 
of letters $x(m, n)$ in the code of $x\in \Si^{\om^2}$ 
(they are enumerated for  increasing values of $m+n$), 
that $h$ is a continuous function from $\Si^{\om^2}$  into 
$((\Si\cup\{A\})\times (\Si \cup\{A\}))^\om$. 

\hs  Notice that we have chosen  $x(1, 1)$ to be  the first letter of the word $\sigma_1$. 
In fact, with slight modifications in the sequel,  
we could have chosen  the letter  $x(1, 1)$ to be the first letter of the word $\sigma_2$. 
\nl  Remark that the above coding of $\om^2$-words resembles the use of the Cantor pairing 
function as it was used to construct the complete sets $P_i$ and $S_i$ in \cite{sw} 
(see also \cite{stac} or \cite[section 3.4]{sta}).

\hs We  now state the following lemmas:

\begin{Lem}\label{lem}
Let $\Si=\{0, 1\}$ and 
$S = \{ x\in \{0, 1\}^{\om^2}~/~ \exists m \exists^\infty n ~x(m, n)=1\}$. Then the set 
$$\mathcal{S}=h(S) \cup  (h(\Si^{\om^2}))^-$$
\noi 
 is a ${\bf \Si^0_3}$-complete subset of 
$((\Si\cup\{A\})\times (\Si \cup\{A\}))^\om$.  
\end{Lem} 

\proo We just sketch the first part of the proof. 
\nl The topological space $\Si^{\om^2}$  is compact and   the  function 
$h$ is continuous and injective. Using these  facts we can easily show that 
$h(S)$ is a ${\bf \Si^0_3}$-subset of   $((\Si\cup\{A\})\times (\Si \cup\{A\}))^\om$. 

\hs On the other side $h(\Si^{\om^2})$ is a closed subset of 
$((\Si\cup\{A\})\times (\Si \cup\{A\}))^\om$. Then its complement 
$$(h(\Si^{\om^2}))^- = ((\Si\cup\{A\})\times (\Si \cup\{A\}))^\om - h(\Si^{\om^2})$$
\noi  is an open (i.e. a ${\bf \Si^0_1}$) subset of 
$((\Si\cup\{A\})\times (\Si \cup\{A\}))^\om$. 
\nl Now $\mathcal{S}=h(S) \cup  (h(\Si^{\om^2}))^-$ is the union of a ${\bf \Si^0_3}$-set 
and of a ${\bf \Si^0_1}$-set therefore it is a ${\bf \Si^0_3}$-set because the class of 
${\bf \Si^0_3}$-subsets of 
$((\Si\cup\{A\})\times (\Si \cup\{A\}))^\om$ is closed under finite union.  

\hs In order to prove that $\mathcal{S}$ is ${\bf \Si^0_3}$-complete it suffices to remark 
that $S=h^{-1}(\mathcal{S})$. This implies that $\mathcal{S}$ is ${\bf \Si^0_3}$-complete 
because $S$ is ${\bf \Si^0_3}$-complete. \ep 

\begin{Lem}
For $\Si$  a finite alphabet, 
$$(h(\Si^{\om^2}))^- = ((\Si\cup\{A\})\times (\Si \cup\{A\}))^\om - h(\Si^{\om^2})$$
\noi is an infinitary rational relation.  
\end{Lem}

\proo  Return to the definition of coding of $\om^2$-words 
over $\Si$ by $\om$-words over the alphabet  $(\Si\cup\{A\}) \times (\Si\cup\{A\})$.   
The code of $x\in \Si^{\om^2}$ was written in the form $(\sigma_1, \sigma_2)$, 
where $\sigma_1$ and $\sigma_2$ are 
$\om$-words over the alphabet $(\Si\cup\{A\})$ in the form:   

\hs $\sigma_1=u_1.A.u_2.A.u_4.A.u_6.A.u_8.A \ldots .A.u_{2n}.A \ldots$

\hs $\sigma_2=A.u_3.A.u_5.A.u_7.A.u_9.A \ldots .A.u_{2n+1}.A \ldots$

\hs where for all integers $i\geq 1$, $u_i \in \Sis$ and $|u_i|=i$. 

\hs It is now easy to see that the complement of the set $h(\Si^{\om^2})$ of codes of 
$\om^2$-words over $\Si$ is the union of the sets $\mathcal{C}_j$ where:

\begin{itemize} 
\ite $\mathcal{C}_1 = \{ (\sigma_1, \sigma_2) ~/~ \sigma_1 , \sigma_2 \in (\Si\cup\{A\})^\om 
\mbox{ and } ( \sigma_1 \in \mathcal{B} \mbox{ or } \sigma_2 \in \mathcal{B} ) \}$
\nl where $\mathcal{B}$ is the set of $\om$-words over $(\Si\cup\{A\})$ having only 
a finite number of letters $A$. 

\ite $\mathcal{C}_2$ is formed by pairs  $(\sigma_1, \sigma_2)$ where 
\nl $\sigma_1$ has not an initial segment in $\Si.A.\Si^2.A$  or 
\nl the first letter of $\sigma_2$ is not an $A$. 

\ite $\mathcal{C}_3$ is formed by pairs  $(\sigma_1, \sigma_2)$ where 
\nl $\sigma_1 = w_1.A.w_2.A.w_3.A.w_4. \ldots A.w_n.A.u.A.z_1 $
\nl $\sigma_2 = w'_1.A.w'_2.A.w'_3.A.w'_4. \ldots A.w'_n.A.v.A.z_2 $

\hs where $n$ is an integer $\geq 1$,  for all $i \leq n$~  $w_i, w'_i \in \Sis$, 
$z_1, z_2 \in (\Si\cup\{A\})^\om$ and 
$$u, v \in \Sis \mbox{  and }  |v|\neq |u|+1$$

\ite $\mathcal{C}_4$ is formed by pairs  $(\sigma_1, \sigma_2)$ where 
\nl $\sigma_1 = w_1.A.w_2.A.w_3.A.w_4. \ldots A.w_n.A.w_{n+1}.A.v.A.z_1 $
\nl $\sigma_2 = w'_1.A.w'_2.A.w'_3.A.w'_4. \ldots A.w'_n.A.u.A.z_2 $

\hs where $n$ is an integer $\geq 1$,  for all $i \leq n$~  $w_i, w'_i \in \Sis$, 
$w_{n+1} \in \Sis$, 
$z_1, z_2 \in (\Si\cup\{A\})^\om$ and 
$$u, v \in \Sis \mbox{  and }  |v|\neq |u|+1$$

\end{itemize}

\noi Each set $\mathcal{C}_j$, $1\leq j\leq 4$, is easily seen to be an infinitary 
rational relation $\subseteq (\Si\cup\{A\})^\om \times (\Si\cup\{A\})^\om$ (the detailed 
proof is left to the reader).  
\nl The class of infinitary rational  relations is closed under finite union;  
this follows from the fact that they are recognized by {\bf non deterministic} 
B\"uchi transducers. Then 

$$(h(\Si^{\om^2}))^- = \bigcup_{1\leq j\leq 4} \mathcal{C}_j$$

\noi is an infinitary rational relation. \ep  

\hs We cannot show directly  that $h(S) \in RAT$ so  we are now 
looking for a rational relation $R \subseteq ((\Si\cup\{A\})\times (\Si \cup\{A\}))^\om$ 
(with $\Si=\{0, 1\}$)  
such that for every  $\om^2$-word $x \in \Si^{\om^2}$, 
$h(x) \in R$ if and only if  $x \in S$. Then  we shall have $S = h^{-1} ( R )$.

\hs We shall first describe the  relation $R$ which  is an \ol~ over the alphabet 
$((\Si\cup\{A\})\times (\Si \cup\{A\}))$.  Every word of $R$ may be seen as a pair 
$y=(y_1, y_2)$ of $\om$-words over the alphabet $\Si\cup\{A\}$ and then $y$ is in $R$ 
if and only if it is in the form

\hs $y_1 = U_k.u.t(1).v_1.A.g_1.t(3).v_2.A.g_2.t(5)  
\ldots A.g_n.t(2n+1).v_{n+1}.A. \ldots$
\nl $y_2 = V_k.u_1.t(2).z_1.A.u_2.t(4).z_2.A. \ldots A.u_n.t(2n).z_n.A \ldots$

\hs where $k$ is an integer $\geq 1$, $U_k, V_k  \in (\Sis.A)^k$, 
$u=\lambda$ or $u \in \Si$, and 
for all integers $i\geq 1$, $t(i)\in \Si$ and  $u_i, v_i, g_i, z_i \in \Sis$ and 
$$|v_i|=|u_i|  ~~~~ \mbox{ and  }  ~~~~|g_i|=|z_i|+1$$ 
and the $\om$-word $t=t(1)t(2)\ldots t(n)\ldots $ is in the \orl~ $\mathcal{A}$ 
given in Example \ref{exa}.  

\begin{Lem}   The  above defined  relation  $R$  satisfies  $S = h^{-1} ( R )$, i.e.:
$$\fa x \in \Si^{\om^2} ~~~~ h(x) \in R  \longleftrightarrow  x \in S$$ 
\end{Lem}

\proo Assume first that  such an $y=(y_1, y_2)\in R$ is the code $h(x)$ of  
an $\om^2$-word $x \in \Si^{\om^2}$.  Then 
$$u.t(1).v_1 = x(2k, 1).x(2k-1, 2)\ldots x(1, 2k)$$
\noi so if $u=\lambda$ then $x(2k,1)=t(1)$ and $|v_1|=2k-1$.  
\nl And if $u\in \Si$ then $x(2k-1,2)=t(1)$ and $|v_1|=2k-2$. 
\nl Next 
$$u_1.t(2).z_1 = x(1, 2k+1).x(2, 2k)\ldots x(2k+1, 1)$$
\noi so if $u=\lambda$ then $|u_1|=|v_1|=2k-1$ thus  $x(2k,2)=t(2)$ and $|z_1|=1$. 
\nl And if $u\in \Si$ then $|u_1|=|v_1|=2k-2$ thus $x(2k-1,3)=t(2)$ and  $|z_1|=2$.  

\hs Moreover $$g_1.t(3).v_2 = x(2k+2, 1).x(2k+1, 2)\ldots x(1, 2k+2)$$
\noi so if $u=\lambda$ then $|g_1|=|z_1|+1=2$ and $t(3)=x(2k, 3)$. 
\nl And if  $u\in \Si$ then $|g_1|=|z_1|+1=3$  and  $t(3)=x(2k-1, 4)$. 

\hs In a similar manner one can  show by induction 
on integers $i$ that if $u=\lambda$ 
 letters $t(i)$ are  successive letters of the $(2k)^{th}$ column of $x$ and if 
$u\in \Si$ letters $t(i)$ are  successive letters of the $(2k-1)^{th}$ column of $x$. 

\hs So assume that for some integer $i\geq 1$: 
\nl if  $u=\lambda$ then $t(2i+1)=x(2k, 2i+1)$, and 
\nl if  $u\in \Si$ then $t(2i+1)=x(2k-1, 2i+2)$. 

\hs We know that 
$$g_i.t(2i+1).v_{i+1} = x(2k+2i, 1).x(2k+2i-1, 2)\ldots x(1, 2k+2i)$$
\noi thus $u=\lambda$ implies that  $|v_{i+1}|=2k-1$ and 
$u\in \Si$  implies that  $|v_{i+1}|=2k-2$. 
\nl But it holds also that
$$u_{i+1}.t(2i+2).z_{i+1}=x(1, 2k+2i+1).x(2, 2k+2i) \ldots x(2k+2i+1, 1)$$
\noi So if $u=\lambda$ then $|u_{i+1}|=|v_{i+1}|=2k-1$ and   $t(2i+2)=x(2k, 2i+2)$ and 
$|z_{i+1}|=2i+1$. 
\nl Anf if $u\in \Si$ then $|u_{i+1}|=|v_{i+1}|=2k-2$ and  $t(2i+2)=x(2k-1, 2i+3)$ and 
$|z_{i+1}|=2i+2$. 

\hs Next 
$$g_{i+1}.t(2i+3).v_{i+2} = x(2k+2i+2, 1).x(2k+2i+1, 2)\ldots x(1, 2k+2i+2)$$
\noi So if $u=\lambda$ then $|g_{i+1}|=|z_{i+1}|+1=2i+2$ and $t(2i+3)=x(2k, 2i+3)$. 
\nl Anf if $u\in \Si$ then $|g_{i+1}|=|z_{i+1}|+1=2i+3$ and $t(2i+3)=x(2k-1, 2i+4)$. 

\hs We have then proved by induction that if  $u=\lambda$,  
\nl $t=t(1).t(2)\ldots t(n)\ldots = x(2k,1).x(2k,2)\ldots x(2k,n)\ldots$ 
\nl  and if $u\in \Si$, 
\nl $t=t(1).t(2)\ldots t(n)\ldots = x(2k-1, 2).x(2k-1, 3)\ldots x(2k-1, n)\ldots$ 

\hs Notice that in this second  case the $\om$-word $t$ begins with the second letter 
$x(2k-1,2)$ of the $(2k-1)^{th}$ column of $x$ and not with the first letter $x(2k-1, 1)$ 
of this column. But this will not change the fact that $t \in \mathcal{A}$ or 
$t \notin \mathcal{A}$ because $\mathcal{A}$ is simply the set of $\om$-words over the 
alphabet $\{0, 1\}$ with infinitely many occurrences of the letter $1$. 

\hs Thus if a code $h(x)$ of  an $\om^2$-word $x \in \Si^{\om^2}$ is in $R$ then 
$x$ has a column in $\mathcal{A}$, i.e. $x\in S$. 
Conversely it is easy to see that every code $h(x)$ of $x\in S$ may be written 
in the above form $(y_1, y_2)$ of a word in $R$.  
Then we have proved that $S = h^{-1} ( R )$.   \ep

\hs  Remark  that the {\it non determinism} of a transducer recognizing $R$ (or of 
a  $2$-tape finite automaton accepting $R$) will be  used 
to guess the integer $k$ and  whether $u=\lambda$ or $u\in \Si$. 
\nl Intuitively, if $\mathcal{T}$ is 
such a transducer recognizing $R$ then, during a successful  
computation accepting the  code $h(x)$ of  
an $\om^2$-word $x$ in $\Si^{\om^2}$, 
 the  {\it non determinism} of $\mathcal{T}$ is used to guess a column of the 
$\om^2$-word $x$ in order  to  simulate on this column the behaviour of a 
 finite B\"uchi automaton accepting  the ${\bf \Pi^0_2 }$-complete \orl~ 
$\mathcal{A}$.

\begin{Lem}
The above defined relation $R$ is an infinitary rational relation.  
\end{Lem}

\proo  This is easy to see from the definitions of $R$ and  of an infinitary rational relation.  
 The infinitary rational relation $R$ is  recognized by the following
B\"uchi transducer 
 $\mathcal{T}=(K, (\{0, 1, A\}), (\{0 ,1, A\}), \Delta, q_0, F)$, 
where 
$$K=\{q_0, q_1, q_2, q_3,  q_1^0, q_1^1, q_2^0, q_2^1\}$$
\noi  is a finite set of states, 
$\{0, 1, A\}=\Si\cup \{A\}$ is the input {\it and} the output alphabet (with $\Si=\{0,1\}$), 
$q_0$ is the initial state,  and $F = \{q_1^1, q_2^1\}$ is the set of 
accepting states. 
Moreover $\Delta \subseteq K \times  
(\Si \cup \{A\})^\star \times (\Si \cup \{A\})^\star \times K$ is the finite  
 set of transitions, containing  the following transitions:

\hs $(q_0, \lambda, a, q_0)$ and $(q_0, a, \lambda, q_0)$,  for all $a\in\Si$, 
\nl  $(q_0, A, A, q_0)$, 
\nl  $(q_0, A, A, q_1)$, 
\nl  $(q_1, u, \lambda,  q_2)$,  for all $u\in \Si\cup \Si^2$, 
\nl  $(q_2, a, b,  q_2)$,  for all $a, b\in \Si$, 
\nl  $(q_2, A, 0, q_1^0)$ and  $(q_2, A, 1, q_1^1)$, 
\nl  $(q, a, \lambda, q_3)$,  for all $a\in \Si$ and $q\in \{q_1^0, q_1^1\}$,
\nl  $(q_3, a, b,  q_3)$,  for all $a, b\in \Si$, 
\nl  $(q_3, 0, A, q_2^0)$ and  $(q_3, 1, A, q_2^1)$, 
\nl  $(q, \lambda, \lambda, q_2)$,  for all  $q\in \{q_2^0, q_2^1\}$. \ep

\hs Return now to the proof of Theorem \ref{si30} and  
consider  the set $\mathcal{R} = R \cup (h(\Si^{\om^2}))^- $. It turns out  that 
$$\mathcal{R} = \mathcal{S} = h(S) \cup (h(\Si^{\om^2}))^- $$
\noi because  $S = h^{-1} ( R )$. But we have proved that $(h(\Si^{\om^2}))^- $ and $R$ 
are infinitary rational relations thus $\mathcal{R} = R \cup (h(\Si^{\om^2}))^- $ is the union 
of two infinitary rational relations hence $\mathcal{R} \in RAT$. 
Lemma \ref{lem} asserts that $\mathcal{R} = \mathcal{S}$ is a
 ${\bf \Si^0_3}$-complete subset of 
$((\Si\cup\{A\})\times (\Si \cup\{A\}))^\om$ and this ends the proof.  \ep 

\begin{Rem}
With a slight modification we could have replaced the set $S$ by the set 
of $\om^2$-words over $\Si$  
having at least one column in a given  ${\bf \Pi^0_2 }$-complete \orl . 
\end{Rem}

\section{${\bf \Pi^0_3}$-complete infinitary  rational relations}

\noi We can now state our next result:

\begin{theorem}\label{pi} There exist some ${\bf \Pi^0_3}$-complete 
infinitary  rational relations. 
\end{theorem}

\proo  We are going to sketch the proof but we cannot give here all details because 
of limited space for this paper. 
\nl As in the last section, we shall use a well known 
example of ${\bf \Pi^0_3}$-complete set which is a 
subset of the topological space $\Si^{\om^2}$ with $\Si=\{0, 1\}$. 

\hs Recall  that the set 
$$P = \{ x\in \{0, 1\}^{\om^2}~/~ \fa m \exists^{<\infty} n ~x(m, n)=1\}$$ 
\noi where $\exists^{<\infty}$ means "there exist only finitely many", 
is a ${\bf \Pi^0_3}$-complete subset of $\{0, 1\}^{\om^2}$, \cite[p. 179]{kec}. 
$P=\{0, 1\}^{\om^2}-S$ so "$P$ is ${\bf \Pi^0_3}$-complete" follows directly from 
"S is ${\bf \Si^0_3}$-complete". 
\nl $P$ is the set of $\om^2$-words 
having  all their columns in the ${\bf \Si^0_2 }$-complete subset $\mathcal{A}^-$ 
of $\{0, 1\}^{\om}$ where $\mathcal{A}$ is the ${\bf \Pi^0_2 }$-complete \orl~
given in Example \ref{exa}. 

\hs We shall use the same coding $h: x\ra h(x)$ for $\om^2$-words over the alphabet 
$\Si=\{0, 1\}$ as in preceding section. 

\begin{Lem}\label{pi1}
$$\mathcal{P} = h(P) \cup (h(\Si^{\om^2}))^-$$
\noi 
 is a ${\bf \Pi^0_3}$-complete subset of 
$((\Si\cup\{A\})\times (\Si \cup\{A\}))^\om$.  
\end{Lem}

\proo It is similar to proof of Lemma \ref{lem}. \ep

\hs We are going to find an infinitary rational relation $R_1$ such that $P=h^{-1}(R_1)$. 
 We define now the relation $R_1$. It  
is an \ol~ over the alphabet 
$((\Si\cup\{A\})\times (\Si \cup\{A\}))$.  Every word of $R_1$ may be seen as a pair 
$y=(y_1, y_2)$ of $\om$-words over the alphabet $\Si\cup\{A\}$ and then $y$ is in $R_1$ 
if and only if it is in the form

\hs $y_1 = U_k.u.t(1).v_1.A.g_1.t(3).v_2.A.g_2.t(5)  
\ldots A.g_n.t(2n+1).v_{n+1}.A. \ldots$
\nl $y_2 = V_k.u_1.t(2).z_1.A.u_2.t(4).z_2.A. \ldots A.u_n.t(2n).z_n.A \ldots$

\hs where $k$ is an integer $\geq 1$, $U_k, V_k  \in (\Sis.A)^k$, $u \in \Sis$, and 
for all integers $i\geq 1$, $t(i)\in \Si$ and  
$$u_i, v_i \in 0^\star \mbox{ and }  g_i, z_i \in \Sis  \mbox{  and   }  $$ 
$$|v_i|=|u_i|  ~~~~ \mbox{ and  }  ~~~~[~~ |g_i|=|z_i|+1 \mbox{ or } |g_i|=|z_i| ~~]$$ 
and there exist infinitely many integers $i$ such that $ |g_i|=|z_i|$. 

\begin{Lem}\label{pi2}
The above defined  relation  $R_1$ satisfies  $P = h^{-1} ( R_1 )$, i.e.:
$$\fa x \in \Si^{\om^2} ~~~~ h(x) \in R_1  \longleftrightarrow  x \in P$$ 
\end{Lem}

\begin{Lem}\label{pi3}
 The above defined  relation $R_1$   is an infinitary rational relation. 
\end{Lem}

\noi Return to the proof of theorem \ref{pi}.   
By Lemma \ref{pi2} the infinitary relation 
$R_1$ satisfies  $P = h^{-1} ( R_1 )$ thus we shall have 
$$\mathcal{P}=  h(P) \cup (h(\Si^{\om^2}))^- = R_1 \cup (h(\Si^{\om^2}))^- $$
\noi But by Lemma \ref{pi3} $R_1$ is rational 
hence $\mathcal{P}$ is  the union of two   infinitary rational relations thus 
$\mathcal{P}\in RAT$ and is ${\bf \Pi^0_3}$-complete by Lemma \ref{pi1}.   \ep

\hs  From Theorems \ref{si30} and   \ref{pi} we can now easily infer the following result:

\begin{Cor} There exists some ${\bf \Delta^0_4}$  (i.e. ${\bf \Si^0_4} \cap {\bf \Pi^0_4}$)   
infinitary  rational relations which are not in $({\bf \Si^0_3} \cup {\bf \Pi^0_3})$. 
\end{Cor}

\noi The question naturally arises whether there exist some  infinitary rational relations 
 $R \subseteq \Sio \times \Ga^\om$  which are ${\bf \Si^0_4}$-complete or 
${\bf \Pi^0_4}$-complete or even higher in the Borel hierarchy. 

\hs {\bf  Acknowledgements.} Thanks to Jean-Pierre Ressayre and  Pierre Simonnet
for useful discussions and to the anonymous referees for useful 
comments on a previous version 
of this paper.

\end{document}